\documentclass[fleqn,usenatbib,useAMS]{mnras}

\usepackage{graphicx}	
\usepackage{amsmath}	
\usepackage{amssymb}	
\usepackage{multicol}        
\usepackage{bm}		
\usepackage{pdflscape}	
\usepackage{xcolor}	

\usepackage[T1]{fontenc}
\usepackage{ae,aecompl}

\usepackage{newtxtext,newtxmath}

\title[A Fast Spinning Polluted White Dwarf]{A Nearby Polluted White Dwarf with a 6.2\,h Spin Period}

\author[J.~Farihi et al.]{Jay Farihi$^1$\thanks{E-mail: j.farihi@ucl.ac.uk},
Akshay Robert$^1$,
Nikolay Walters$^1$
\\
$^1$Department of Physics and Astronomy, University College London, London WC1E 6BT, UK\\
}

\begin{document}


\maketitle




\begin{abstract}
This letter reports the first detection of a periodic light curve whose modulation is unambiguously due to rotation in a polluted white dwarf.  {\em TESS} observations of WD\,2138--332, at a distance of 16.1\,pc, reveal a 0.39\,per cent amplitude modulation with a 6.19\,h period.  While this rotation is relatively rapid for isolated white dwarfs, it falls within the range of spin periods common to those with detectable magnetic fields, where WD\,2138--332 is notably both metal-rich and weakly magnetic.  Within the local 20\,pc volume of white dwarfs, multi-sector  {\em TESS} data find no significant periodicities among the remaining 16 polluted objects (five of which are also magnetic), whereas six of 23 magnetic and metal-free targets have light curves consistent with rotation periods between 0.7 and 35\,h (three of which are new discoveries).  This indicates the variable light curve of WD\,2138--332 is primarily a result of magnetism, as opposed to an inhomogeneous distribution of metals.  From 13 magnetic and metallic degenerates with acceptable  {\em TESS} data, a single detection of periodicity suggests that polluted white dwarfs are not rotating as rapidly as their magnetic counterparts, and planet ingestion is thus unlikely to be a significant channel for rapid rotation.
\end{abstract}


\section{Introduction}

Metal pollution and magnetism are each independently common in white dwarfs.  The ubiquitous metal-enrichment observed in white dwarfs is well-established at roughly 20 to 30\,per cent \citep{zuckerman2003,zuckerman2010}, and sufficiently sensitive spectroscopy can extend to unbiased samples beyond 100\,pc \citep{koester2014}.  In contrast, the intrinsic occurrence rate of magnetism in white dwarfs suffered from observational bias favoring more distant, luminous targets with fields above 1\,MG \citep{ferrario2015}.  Recently however, based on spectropolarimetric observations of the entire 20\,pc volume of white dwarfs (e.g.\ \citealt{landstreet2019,bagnulo2020}), it is now known that around 20\,per cent of white dwarfs are magnetic, in agreement with earlier estimates (e.g.\ \citealt{kawka2007}).  In this pioneering effort, one of several key insights is that magnetic fields are more frequently present and detectable around cooler white dwarfs \citep{bagnulo2021}.

Importantly, there is a bias {\em against} detecting magnetism in cooler white dwarfs, owing to the weakening and eventual disappearance of helium and hydrogen spectral lines, and thus the fact that fields appear relatively late is robust.  The presence of metals in cool white dwarf thus provide spectral lines which are sensitive to the presence of a magnetic field, whereas metal-free counterparts are often featureless or may exhibit Swan bands that are relatively insensitive tracers of polarization \citep{bagnulo2019}.

The first mention of a possible, speculative connection between metals and magnetism was based on the discovery of apparently double Ca\,{\sc ii} H and K lines in the nearby white dwarf G77-50, where this was initially interpreted as duplicity \citep{zuckerman2003}.  Detailed follow up spectroscopy revealed more complex splitting consistent with a magnetic field of approximately 120\,kG \citep{farihi2011b}.  At the time, it was suggested that if a common envelope in binary stars might be responsible for strong magnetic field generation \citep{tout2008}, then perhaps giant planet engulfment might lead to similar outcomes, but with weaker results.  However, it was quickly noted that there is a dearth of sufficient progenitor systems (i.e.\ hot Jupiters), at least around solar-type stars \citep{kawka2011}.  Possibly more relevant, radial velocity studies continue to suggest that giant planets, orbiting sufficiently close to be later engulfed, are more common around the A-type progenitors of white dwarfs \citep{bowler2010,ghezzi2018}.

The literature has continued to discuss a possible, observed correlation between white dwarf magnetism and photospheric metals \citep{kawka2014,kawka2019,bagnulo2019}, and there are recent speculations that planetary accretion may enhance spin rates \citep{schreiber2021}.  The apparent correlation is now understood as a consequence of white dwarf cooling age, combined with a bias against detecting fields in cool white dwarfs without metal lines \citep{bagnulo2021}.  Nevertheless, the correlation with cooling age implies that if an older white dwarf has spectral lines, then it has around a 20\,per cent chance to have a detectable magnetic field, and this applies equally to polluted and non-polluted white dwarfs.

\begin{figure*}
\includegraphics[width=\columnwidth]{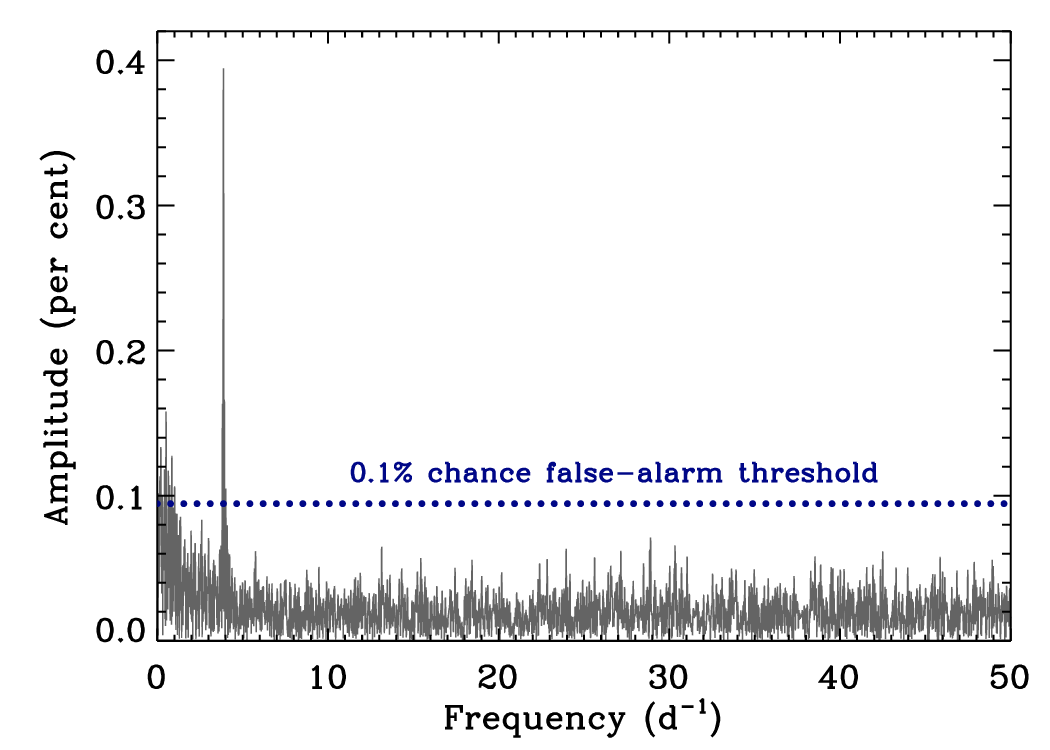}
\includegraphics[width=\columnwidth]{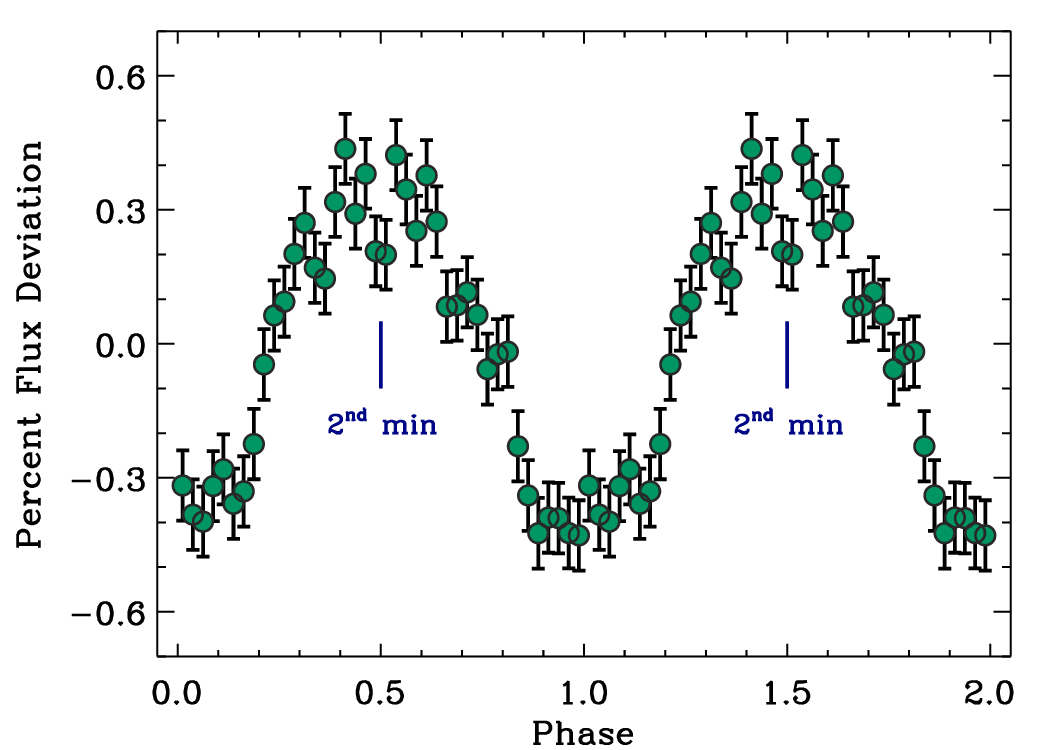}
\vskip 0pt
\caption{{\em Left}:  A periodogram of WD\,2138--332 based on Sector 68 {\em TESS} observations, where the amplitudes are plotted in grey, and a 0.1\,per cent probability, false-alarm threshold (coincidentally at an amplitude of 0.095\,per cent) is plotted with a blue dotted line.  There is a single, strong peak frequency at 3.87\,d$^{-1}$ (6.2\,h).  {\em Right}:  The phased {\em TESS} light curve of WD\,2138--332, with 40 equally-sized phase bins, where the fluxes and errors are weighted means of approximately 370 measurements each.  A modest yet clear secondary minimum is detected 180\,\degr\ from the primary minimum, and consistent with two spots in a dipolar configuration.
\label{pgram_lc}}
\end{figure*}

Despite a few dozen examples of magnetic and metallic white dwarfs discovered over more than two decades \citep{reid2001,dufour2006}, until now none were known to have a rotationally modulated light curve.  In fact, among polluted white dwarfs that are sufficiently cool that radiative forces are negligible, no star is known to exhibit rotational modulation via spectroscopy or photometry.  In hotter white dwarfs where radiative levitation of heavy elements can play a role, there does not appear to be a correlation with light curve modulation and photospheric metals \citep{hallakoun2018,hoard2018}.  And while there are observed periodicities in the optical, it is unclear if the source is rotational modulation or has a circumstellar origin \citep{wilson2020}.

This letter reports unambiguous rotational modulation in the optical light curve of the $T_{\rm eff}=6900$\,K, DZ spectral type white dwarf WD\,2138--332 via space- and ground-based photometry.  The modulation is broadly sinusoidal but with an amplitude of less than 0.6\,per cent at wavelengths longer than 4000\,\AA, and consistent with a stellar spin period of 6.2\,h.  In Section~2 the data and observations are described, with a periodogram analysis and phase-folded light curve presented in Section~3.  The detected light curve amplitude of WD\,2138--332 is compared with detections and upper limits for similar stars in Section~4, followed by a short discussion.

\section{Observations and data}

At $d=16.1$\,pc and with $G=14.5$\,mag, WD\,2138--332 (= NLTT\,51844) is one of the nearest and brightest white dwarfs, and was not discovered until a relatively recent, dedicated search in the southern hemisphere \citep{subasavage2007}.  The star exhibits a classical DZ spectrum with strong lines of Ca\,{\sc ii} H \& K, as well as other lines of Mg and Fe, even at low resolution.  A model atmosphere fit based on spectroscopy, photometry, and {\em Gaia} parallax yields $T_{\rm eff}\approx6900$\,K and $M=0.60$\,M$_{\odot}$, with around 10\,per cent and 20\,per cent errors, respectively, and corresponding to a nominal cooling age of 1.7\,Gyr \citep{coutu2019}.  The magnetic field was recently detected with spectropolarimetry, and is one of the weakest field strengths measured at approximately 50\,kG, yet robust with multiple detections \citep{bagnulo2019,bagnulo2021}.

{\em Transiting Exoplanet Survey Satellite} \citep[{\em TESS};][]{ricker2015} data for WD\,2138--332 (TIC\,204440456, $T=14.2$\,mag) were retrieved from the MAST archive\footnote{https://archive.stsci.edu}, and Sector 68 {\sc pdcsap} light curves with 120\,s cadence were downloaded \citep{jenkins2016}, where 96\,per cent of the total flux in the extracted aperture is attributable to the target white dwarf.  The Sector 68 data span the dates from 2023 Jul 29 to Aug 25.  There are no other useful {\em TESS} observations for this star, where it was imaged in Sector 28, but too close to the detector edge for data to be extracted.  Minimal cleaning of the light curve was performed, including outlier rejection and removal of NaN entries, and the time stamps were adjusted to BJD = TBJD + 2457000.  The {\em TESS} filter samples wavelengths in the range 6000--10\,000\,\AA, similar to Cousins $I$-band in central wavelength but significantly broader \citep{ricker2015}.

Antecedent and subsequent observations of the white dwarf were obtained using ULTRACAM \citep{dhillon2007} on the NTT at La Silla Observatory on 2021 Aug 20 and 2023 Sep 15, resulting in light curve durations of approximately 2.0 and 4.0\,h, respectively.  The instrument is a triple-beam, frame transfer imaging camera, and images were taken in high throughput $ugr$ bandpass filters, with 3.2\,s exposures on both observing dates.  Images in the blue channel were co-added every three frames to increase signal-to-noise (S/N), but otherwise all three channel images were taken simultaneously.

The raw frames were bias corrected and flat fielded using sky flats taken during evening twilight each night.  Differential photometry was performed on the resulting images using custom-built software\footnote{https://github.com/HiPERCAM/hipercam}.  Owing to the fact that WD\,2138--332 is a relatively bright white dwarf with $(G_{\rm BP},G,G_{\rm RP})= (14.61,14.45,14.19)$\,mag, there was only a single field star that was suitably bright as a comparison source, {\em Gaia}\,DR3\,6592315860631132800 with $(G_{\rm BP},G,G_{\rm RP})= (14.64,14.25,13.69)$\,mag.  The typical photometric S/N per saved (co-added or single) image was 130, 290, and 280 in $ugr$ respectively for the target star, and 90, 280, and 300 for the comparison star.  

ULTRACAM light curves were constructed by dividing the science target flux by that of the comparison star, and with errors propagated as the quadrature sum of the fractional flux errors of both stars measured.  Thus the resulting light curves have formal errors that are typically 1\,per cent or better.  Time stamps were converted to BJD following \citet{eastman2010}.

\begin{figure}
\includegraphics[width=\columnwidth]{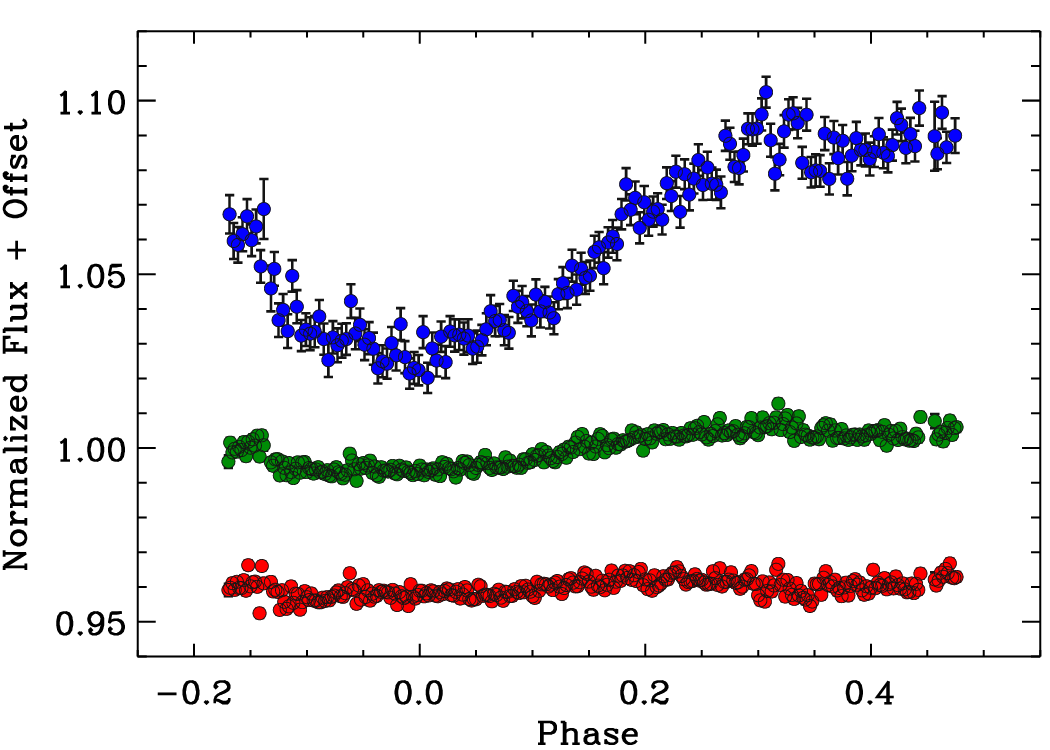}
\vskip 0pt
\caption{Phased light curves of WD\,2138--332 from ULTRACAM.  The blue data points and errors correspond to $u$-band photometry, and similarly the green and red symbols are $g$- and $r$-band data, respectively.  The plotted data are the most recent observing run from 2023 Sep, where the 2021 Aug data offer no additional phase coverage and are thus not plotted.  The bin sizes are 0.002 (500  bins) and 0.004 (250 bins), respectively for the $u$-band light curve, and both the $g$- and $r$-band light curves.  The variability amplitude in the $u$-band is remarkable, and the corresponding light curve shows partial coverage of the secondary minimum seen in the {\em TESS} data.
\label{ucam}}
\end{figure}

\section{Analysis and results}

In this section, the {\em TESS} and ULTRACAM light curves are analyzed and presented, and WD\,2138--332 is placed into the wider context of the polluted and magnetic white dwarfs within 20\,pc, and beyond.

\subsection{Light curves}

Light curve analysis was carried out using {\sc period04} \citep{lenz2005}, where Fourier power spectra were constructed, and fitted parameter uncertainties calculated using Monte Carlo methods.  These simulations were performed with the frequency and phase uncoupled, and run 1000 times with Gaussian noise added to each point in the fitted light curve resulting from the periodogram analysis.  Figure~\ref{pgram_lc} displays the periodogram based on using only the {\em TESS} Sector 68 data, which yield a robust peak signal at 3.87\,d$^{-1}$ (6.2\,h), and an error of 200\,ppm.  To reduce the uncertainty in the period, a second Fourier analysis was performed using both {\em TESS} and ULTRACAM $g+r$ light curves, generating a refined period but with an uncertainty of 5\,ppm.  The adopted ephemeris is:\\

\noindent
BJD$_{\rm TDB} = 2460203.519(1) + 0.258001(1)\,E$\\

\noindent
The {\em TESS} data were phase folded on this period and are plotted in Figure~\ref{pgram_lc}.  The amplitude of the variation in the {\em TESS} bandpass is 0.39\,per cent.

The 2023 ULTRACAM light curves are shown in Figure~\ref{ucam}, and were fitted with sine waves using the period determined above, and the resulting amplitudes and errors from these fits are listed in Table~\ref{lcamp}.  In terms of detecting low-amplitude photometric variations in white dwarfs, it is notable that the $u$-band variability is an order of magnitude larger than that observed in {\em TESS}.  Both sets of light curves exhibit partial or total coverage of a secondary photometric minimum (e.g.\ Figure~\ref{pgram_lc}), supporting the presence of two surface spots.

\begin{table}
\begin{center}
\caption{Multi-wavelength variability amplitudes in per cent flux\label{lcamp}.}
\begin{tabular}{@{}rcl@{}}

\hline

Band		&$\uplambda_{\rm c}$	&Amplitude\\
		&(\AA)			&(per cent)\\

\hline

$u$		&3600			&$3.19 \pm 0.06$\\
$g$		&4700			&$0.57 \pm 0.01$\\
$r$		&6200			&$0.18 \pm 0.01$\\
$T$		&7900			&$0.39 \pm 0.02$\\


\hline

\end{tabular}
\end{center}
\end{table}

\subsection{Magnetic and metallic stars within 20\,pc and beyond}

To place the discovery of small-amplitude variability in the {\em TESS} light curve of WD\,2138--332 into context, a similar search was carried out for the 20\,pc sample of white dwarfs with either photospheric metals or magnetic fields, as well as all available stars with both properties \citep{schreiber2021}.  For each target, available sectors of {\em TESS} {\sc pdcsap} light curve data were analyzed to either detect or confirm known variability, or in the case where no variability is present, to establish a corresponding sensitivity.  To determine a limit on any variation that should have been detected if present, a 0.1\,per cent probability, false-alarm amplitude was established by following standard methodology \citep{hermes2015}.

\begin{figure}
\includegraphics[width=\columnwidth]{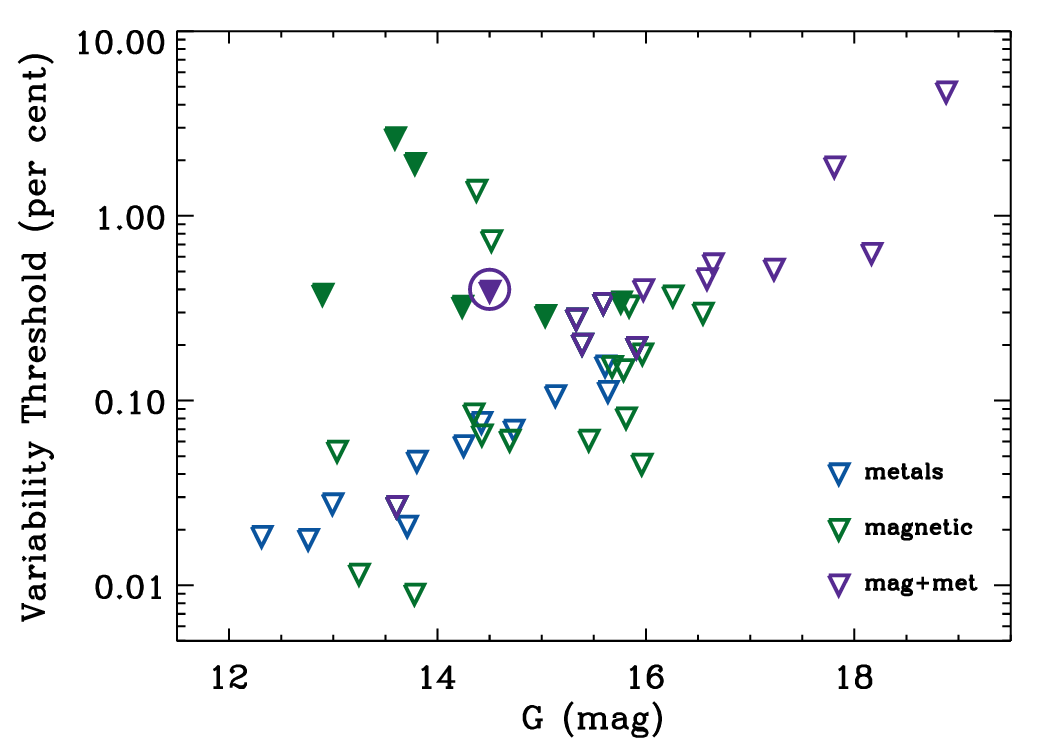}
\vskip 0pt
\caption{{\em TESS} threshold and detected variability amplitudes for metal-rich or magnetic white dwarfs in the 20\,pc sample, and for those with both properties known outside this volume.  The open symbols are 0.1\,per cent probability, false-alarm thresholds for those sources with no significant periodogram peaks, while the filled symbols are actual detected variation amplitudes for those stars with unambiguous periodicities.  Despite issues of crowding within the relatively large {\em TESS} camera pixels, and light curve contamination from a few variable neighboring sources that augments their false-alarm thresholds, the plot demonstrates excellent sensitivity in numerous cases, where WD\,2138--332 (circled) is the only metal-rich white dwarf to have a detected rotation period.
\label{thresh}}
\end{figure}

In Figure~\ref{thresh} are plotted the resulting variability thresholds and actual detected amplitudes among the magnetic and metallic white dwarfs, as a function of source brightness.  Not all target sources have {\sc pdcsap} light curves, and those which do often have unfavorable data owing to crowding next to bright sources.  Furthermore, a small number of sources have light curves that are contaminated by variable sources within the photometric aperture (see Table~\ref{newvar}).  Yet there are numerous sources with sufficient quality data such that variability should have been detectable at the 0.1\,per cent level, and even lower in roughly one dozen cases.

Table~\ref{newvar} lists three magnetic white dwarfs reported for the first time to be photometrically variable.  Also included in the table are {\em TESS} periods for three additional magnetic white dwarfs whose rotational periods had previously been determined using time-series spectropolarimetry.  Thus all six variable sources are consistent with stellar rotation.  Among the new findings is a 44.2\,min rotation period for WD\,0011--134, which has been known to be magnetic and variable for over 25 years, but whose period was never determined until now \citep{bergeron1992,putney1997}.

It has been claimed that the $G=20.0$\,mag, polluted white dwarf DES\,J214756.46--403529.3 is weakly magnetic with a spin period of 13\,h \citep{elms2022}.  Here, an independent analysis finds no significant flux from this $T=19.1$\,mag star in the relevant {\em TESS} Sectors 1 and 28.  Using PSF-subtracted, full-frame images at the {\em Gaia} DR3 position \citep[TGLC;][]{han2023}, aperture fluxes have median and scatter $3.5\pm2.4$\,e$^-$\,s$^{-1}$ in both sectors, comparable with variations in the sky after background subtraction.  {\em Gaia}\,DR3\,6584429922718418176 is 39\,arcsec ($<2$~pixels) distant, and $10\times$ brighter than the white dwarf with TGLC aperture fluxes $38\pm2$\,e$^-$\,s$^{-1}$ in both sectors.  These light curves exhibit a 13\,h periodicity with amplitude 2.2 and 1.1\,per cent in Sectors 1 and 28, respectively (cf.\ 1.5\,per cent reported by \citealt{elms2022}).  Thus, it is challenging to ascribe this periodic signal to the white dwarf.

%


%
%
%
%

%
%
%
%

\section{Discussion}

Only the stars known to be magnetic are detected as variable, and none of these besides WD\,2138--332 have photospheric metals.  This is a strong indication that the photometric variability in WD\,2138--332 originates in magnetism and not with surface patches associated with its metal-enriched atmosphere.  Given that magnetic white dwarfs are commonly photometric variables -- just over 20\,per cent (6 of 29) within the 20\,pc volume -- then one would expect a similar fraction of polluted white dwarfs {\em with detected magnetic fields} to be variable.

There are only 13 magnetic and metallic white dwarfs with sufficient {\em TESS} data to ask if the variable fraction is similar to magnetic stars in general.  In this case, only one of 13 (WD\,2038--332) is variable, whereas a nominal expectation would be at least two.  Owing to small number statistics, these two fractions are insufficiently distinct for any strong conclusions to be drawn.  However, one possibility is that a similar fraction of magnetic white dwarfs with metals have surface spots, but that their periods are too long to be detected in {\em TESS} (cf.\ the suggestion of a 29\,d period in the Zeeman splitting of the Ca\,{\sc ii} lines of G77-50; \citealt{farihi2011b}).  The {\sc pdcsap} data are de-trended, and thus likely unreliable for periods longer than a few days.  Furthermore, if any longer period variations have photometric amplitudes below 1\,per cent as for WD\,2138--332, then detection will remain a challenge, as large-scale photometric surveys often lack the required sensitivity.

Figure~\ref{ucam} suggests that a large-scale survey conducted in the $u$-band might be particularly fruitful for identifying periods in magnetic white dwarfs.  In the case that photometric variability amplitudes are of an order of magnitude more prominent in the bluest optical bandpasses such as $u$, then {\em TESS} may not be ideally suited if the bulk of potentially variable sources have sufficiently low amplitudes.  Recently, it was shown that for a sub-group of variable magnetic white dwarfs, the photometric amplitudes were also strongest in the $u$-band \citep{farihi2023}, suggesting future synoptic surveys employing this bandpass will be fruitful, e.g.\ BlackGem \citep{groot2022}.

Lastly, there is the question of the rotation period of WD\,2138--332 and its origin.  Over several Gyr, polluted white dwarfs may accrete up to a Pluto or even a Lunar mass of material \citep{farihi2012b,swan2023}, but this is unlikely to approach the amount needed to appreciably spin up a white dwarf.  Based on modeling of planet ingestion, gas dwarf or giant planets are likely necessary to attain spin periods less than one day \citep{stephan2020}.  While this may be the case for WD\,2138--332, it appears unlikely to be common for polluted white dwarfs in general.


\begin{table}
\begin{center}
\caption{{\em TESS} spin periods for magnetic white dwarfs within 20\,pc\label{newvar}.}
\begin{tabular}{@{}clcl@{}}

\hline

Star			&Period			&Amplitude		&References\\
			&(d)				&(per cent)		&\\

\hline

WD\,0009+501 	&0.16707(5) 		&$0.31\pm0.02$	&1\\
WD\,0011--134 	&0.030679(3)		&$0.34\pm0.05$	&1\\
WD\,0011--721 	&0.588905(3)		&$0.28\pm0.01$	&1\\
WD\,0912+536 	&1.331045(3)		&$1.94\pm0.01$	&2,1\\
WD\,1953--011	&1.4427(3)		&$2.65\pm0.02$	&3,1\\
WD\,2138--332	&0.25822(6)		&$0.39\pm0.02$	&1\\
WD\,2359--434	&0.11229383(2) 	&$0.375\pm0.003$	&4,1\\

\hline

\end{tabular}
\end{center}
{\em Refs}: (1) This work; (2) \citet{angel1972}; (3) \citet{valyavin2008}; (4) \citet{gary2013}.

{\em Note}: The magnetic white dwarfs WD\,1009--184, WD\,1036--204, and WD\,2153--512 all have variable {\em TESS} light curves, but the sources are attributed to neighboring stars using {\sc tess localize} \citep{higgins2023}.  The period derived for WD\,2138--332 based on {\em TESS} data alone (baseline of 25\,d) differs by 3.6$\upsigma$ from the adopted period that is based on the inclusion of ULTRACAM light curves (baseline of 756\,d).

\end{table}


\section*{Acknowledgements} 

J.~Farihi is grateful to the staff at La Silla Observatory for their support while carrying out observations, and the authors collectively thank the reviewer for their report.  This research was supported by the Munich Institute for Astro, Particle, and BioPhysics which is funded by the Deutsche Forschungsgemeinschaft under Germany's Excellence Strategy EXC\,2094 – 390783311.  A.~Robert and N.~Walters have been supported by UK STFC studentships. For the purpose of open access, the authors have applied a creative commons attribution (CC BY) license to any author accepted manuscript version arising. This paper includes data collected by the {\em TESS} mission, which is funded by the NASA Explorer Program.

\section*{Data Availability}
ULTRACAM data are available on reasonable request to the instrument team, while {\em TESS} data are available through the Mikulski Archive for Space Telescopes.

\bibliographystyle{mnras}

\bibliography{../../references}

\begin{thebibliography}{}
\makeatletter
\relax
\def\mn@urlcharsother{\let\do\@makeother \do\$\do\&\do\#\do\^\do\_\do\%\do\~}
\def\mn@doi{\begingroup\mn@urlcharsother \@ifnextchar [ {\mn@doi@}
  {\mn@doi@[]}}
\def\mn@doi@[#1]#2{\def\@tempa{#1}\ifx\@tempa\@empty \href
  {http://dx.doi.org/#2} {doi:#2}\else \href {http://dx.doi.org/#2} {#1}\fi
  \endgroup}
\def\mn@eprint#1#2{\mn@eprint@#1:#2::\@nil}
\def\mn@eprint@arXiv#1{\href {http://arxiv.org/abs/#1} {{\tt arXiv:#1}}}
\def\mn@eprint@dblp#1{\href {http://dblp.uni-trier.de/rec/bibtex/#1.xml}
  {dblp:#1}}
\def\mn@eprint@#1:#2:#3:#4\@nil{\def\@tempa {#1}\def\@tempb {#2}\def\@tempc
  {#3}\ifx \@tempc \@empty \let \@tempc \@tempb \let \@tempb \@tempa \fi \ifx
  \@tempb \@empty \def\@tempb {arXiv}\fi \@ifundefined
  {mn@eprint@\@tempb}{\@tempb:\@tempc}{\expandafter \expandafter \csname
  mn@eprint@\@tempb\endcsname \expandafter{\@tempc}}}

\bibitem[\protect\citeauthoryear{{Angel}, {Illing}  \& {Landstreet}}{{Angel}
  et~al.}{1972}]{angel1972}
{Angel} J.~R.~P.,  {Illing} R.~M.~E.,   {Landstreet} J.~D.,  1972, \mn@doi
  [\apjl] {10.1086/180990}, \href
  {https://ui.adsabs.harvard.edu/abs/1972ApJ...175L..85A} {175, L85}

\bibitem[\protect\citeauthoryear{{Bagnulo} \& {Landstreet}}{{Bagnulo} \&
  {Landstreet}}{2019}]{bagnulo2019}
{Bagnulo} S.,  {Landstreet} J.~D.,  2019, \mn@doi [\aap]
  {10.1051/0004-6361/201936068}, \href
  {https://ui.adsabs.harvard.edu/abs/2019A&A...630A..65B} {630, A65}

\bibitem[\protect\citeauthoryear{{Bagnulo} \& {Landstreet}}{{Bagnulo} \&
  {Landstreet}}{2020}]{bagnulo2020}
{Bagnulo} S.,  {Landstreet} J.~D.,  2020, \mn@doi [\aap]
  {10.1051/0004-6361/202038565}, \href
  {https://ui.adsabs.harvard.edu/abs/2020A&A...643A.134B} {643, A134}

\bibitem[\protect\citeauthoryear{{Bagnulo} \& {Landstreet}}{{Bagnulo} \&
  {Landstreet}}{2021}]{bagnulo2021}
{Bagnulo} S.,  {Landstreet} J.~D.,  2021, \mn@doi [\mnras]
  {10.1093/mnras/stab2046}, \href
  {https://ui.adsabs.harvard.edu/abs/2021MNRAS.507.5902B} {507, 5902}

\bibitem[\protect\citeauthoryear{{Bergeron}, {Ruiz}  \& {Leggett}}{{Bergeron}
  et~al.}{1992}]{bergeron1992}
{Bergeron} P.,  {Ruiz} M.-T.,   {Leggett} S.~K.,  1992, \mn@doi [\apj]
  {10.1086/171997}, \href
  {https://ui.adsabs.harvard.edu/abs/1992ApJ...400..315B} {400, 315}

\bibitem[\protect\citeauthoryear{{Bowler} et~al.,}{{Bowler}
  et~al.}{2010}]{bowler2010}
{Bowler} B.~P.,  et~al., 2010, \mn@doi [\apj] {10.1088/0004-637X/709/1/396},
  \href {https://ui.adsabs.harvard.edu/abs/2010ApJ...709..396B} {709, 396}

\bibitem[\protect\citeauthoryear{{Coutu}, {Dufour}, {Bergeron}, {Blouin},
  {Loranger}, {Allard}  \& {Dunlap}}{{Coutu} et~al.}{2019}]{coutu2019}
{Coutu} S.,  {Dufour} P.,  {Bergeron} P.,  {Blouin} S.,  {Loranger} E.,
  {Allard} N.~F.,   {Dunlap} B.~H.,  2019, \mn@doi [\apj]
  {10.3847/1538-4357/ab46b9}, \href
  {https://ui.adsabs.harvard.edu/abs/2019ApJ...885...74C} {885, 74}

\bibitem[\protect\citeauthoryear{{Dhillon} et~al.,}{{Dhillon}
  et~al.}{2007}]{dhillon2007}
{Dhillon} V.~S.,  et~al., 2007, \mn@doi [\mnras]
  {10.1111/j.1365-2966.2007.11881.x}, \href
  {https://ui.adsabs.harvard.edu/abs/2007MNRAS.378..825D} {378, 825}

\bibitem[\protect\citeauthoryear{{Dufour}, {Bergeron}, {Schmidt}, {Liebert},
  {Harris}, {Knapp}, {Anderson}  \& {Schneider}}{{Dufour}
  et~al.}{2006}]{dufour2006}
{Dufour} P.,  {Bergeron} P.,  {Schmidt} G.~D.,  {Liebert} J.,  {Harris} H.~C.,
  {Knapp} G.~R.,  {Anderson} S.~F.,   {Schneider} D.~P.,  2006, \mn@doi [\apj]
  {10.1086/508144}, \href
  {https://ui.adsabs.harvard.edu/abs/2006ApJ...651.1112D} {651, 1112}

\bibitem[\protect\citeauthoryear{{Eastman}, {Siverd}  \& {Gaudi}}{{Eastman}
  et~al.}{2010}]{eastman2010}
{Eastman} J.,  {Siverd} R.,   {Gaudi} B.~S.,  2010, \mn@doi [\pasp]
  {10.1086/655938}, \href
  {https://ui.adsabs.harvard.edu/abs/2010PASP..122..935E} {122, 935}

\bibitem[\protect\citeauthoryear{{Elms}, {Tremblay}, {G{\"a}nsicke}, {Koester},
  {Hollands}, {Gentile Fusillo}, {Cunningham}  \& {Apps}}{{Elms}
  et~al.}{2022}]{elms2022}
{Elms} A.~K.,  {Tremblay} P.-E.,  {G{\"a}nsicke} B.~T.,  {Koester} D.,
  {Hollands} M.~A.,  {Gentile Fusillo} N.~P.,  {Cunningham} T.,   {Apps} K.,
  2022, \mn@doi [\mnras] {10.1093/mnras/stac2908}, \href
  {https://ui.adsabs.harvard.edu/abs/2022MNRAS.517.4557E} {517, 4557}

\bibitem[\protect\citeauthoryear{{Farihi}, {Dufour}, {Napiwotzki}  \&
  {Koester}}{{Farihi} et~al.}{2011}]{farihi2011b}
{Farihi} J.,  {Dufour} P.,  {Napiwotzki} R.,   {Koester} D.,  2011, \mn@doi
  [\mnras] {10.1111/j.1365-2966.2011.18325.x}, \href
  {https://ui.adsabs.harvard.edu/abs/2011MNRAS.413.2559F} {413, 2559}

\bibitem[\protect\citeauthoryear{{Farihi}, {G{\"a}nsicke}, {Wyatt}, {Pringle}
  \& {King}}{{Farihi} et~al.}{2012}]{farihi2012b}
{Farihi} J.,  {G{\"a}nsicke} B.~T.,  {Wyatt} M.~C.~{Girven} J.,  {Pringle}
  J.~E.,   {King} A.~R.,  2012, \mn@doi [\mnras]
  {10.1111/j.1365-2966.2012.21215.x}, \href
  {https://ui.adsabs.harvard.edu/abs/2012MNRAS.424..464F} {424, 464}

\bibitem[\protect\citeauthoryear{{Farihi}, {Hermes}, {Littlefair}, {Howarth},
  {Walters}  \& {Parsons}}{{Farihi} et~al.}{2023}]{farihi2023}
{Farihi} J.,  {Hermes} J.~J.,  {Littlefair} S.~P.,  {Howarth} I.~D.,  {Walters}
  N.,   {Parsons} S.~G.,  2023, \mn@doi [\mnras] {10.1093/mnras/stad2184},
  \href {https://ui.adsabs.harvard.edu/abs/2023MNRAS.525.1097F} {525, 1097}

\bibitem[\protect\citeauthoryear{{Ferrario}, {de Martino}  \&
  {G{\"a}nsicke}}{{Ferrario} et~al.}{2015}]{ferrario2015}
{Ferrario} L.,  {de Martino} D.,   {G{\"a}nsicke} B.~T.,  2015, \mn@doi [\ssr]
  {10.1007/s11214-015-0152-0}, \href
  {https://ui.adsabs.harvard.edu/abs/2015SSRv..191..111F} {191, 111}

\bibitem[\protect\citeauthoryear{{Gary}, {Tan}, {Curtis}, {Tristram}  \&
  {Fukui}}{{Gary} et~al.}{2013}]{gary2013}
{Gary} B.~L.,  {Tan} T.~G.,  {Curtis} I.,  {Tristram} P.~J.,   {Fukui} A.,
  2013, Society for Astronomical Sciences Annual Symposium, \href
  {https://ui.adsabs.harvard.edu/abs/2013SASS...32...71G} {32, 71}

\bibitem[\protect\citeauthoryear{{Ghezzi}, {Montet}  \& {Johnson}}{{Ghezzi}
  et~al.}{2018}]{ghezzi2018}
{Ghezzi} L.,  {Montet} B.~T.,   {Johnson} J.~A.,  2018, \mn@doi [\apj]
  {10.3847/1538-4357/aac37c}, \href
  {https://ui.adsabs.harvard.edu/abs/2018ApJ...860..109G} {860, 109}

\bibitem[\protect\citeauthoryear{{Groot} et~al.,}{{Groot}
  et~al.}{2022}]{groot2022}
{Groot} P.~J.,  et~al., 2022, in {Marshall} H.~K.,  {Spyromilio} J.,   {Usuda}
  T.,  eds,  Society of Photo-Optical Instrumentation Engineers (SPIE)
  Conference Series Vol. 12182, Ground-based and Airborne Telescopes IX. p.
  121821V, \mn@doi{10.1117/12.2630160}

\bibitem[\protect\citeauthoryear{{Hallakoun} et~al.,}{{Hallakoun}
  et~al.}{2018}]{hallakoun2018}
{Hallakoun} N.,  et~al., 2018, \mn@doi [\mnras] {10.1093/mnras/sty257}, \href
  {https://ui.adsabs.harvard.edu/abs/2018MNRAS.476..933H} {476, 933}

\bibitem[\protect\citeauthoryear{{Han} \& {Brandt}}{{Han} \&
  {Brandt}}{2023}]{han2023}
{Han} T.,  {Brandt} T.~D.,  2023, \mn@doi [\aj] {10.3847/1538-3881/acaaa7},
  \href {https://ui.adsabs.harvard.edu/abs/2023AJ....165...71H} {165, 71}

\bibitem[\protect\citeauthoryear{{Hermes} et~al.,}{{Hermes}
  et~al.}{2015}]{hermes2015}
{Hermes} J.~J.,  et~al., 2015, \mn@doi [\mnras] {10.1093/mnras/stv1053}, \href
  {https://ui.adsabs.harvard.edu/abs/2015MNRAS.451.1701H} {451, 1701}

\bibitem[\protect\citeauthoryear{{Higgins} \& {Bell}}{{Higgins} \&
  {Bell}}{2023}]{higgins2023}
{Higgins} M.~E.,  {Bell} K.~J.,  2023, \mn@doi [\aj]
  {10.3847/1538-3881/acb20c}, \href
  {https://ui.adsabs.harvard.edu/abs/2023AJ....165..141H} {165, 141}

\bibitem[\protect\citeauthoryear{{Hoard}, {Howell}, {Roettenbacher}, {Ely},
  {Debes}  \& {Harmon}}{{Hoard} et~al.}{2018}]{hoard2018}
{Hoard} D.~W.,  {Howell} S.~B.,  {Roettenbacher} R.~M.,  {Ely} J.,  {Debes}
  J.~H.,   {Harmon} R.~O.,  2018, \mn@doi [\aj] {10.3847/1538-3881/aad238},
  \href {https://ui.adsabs.harvard.edu/abs/2018AJ....156..119H} {156, 119}

\bibitem[\protect\citeauthoryear{{Jenkins} et~al.,}{{Jenkins}
  et~al.}{2016}]{jenkins2016}
{Jenkins} J.~M.,  et~al., 2016, in \procspie. p. 99133E,
  \mn@doi{10.1117/12.2233418}

\bibitem[\protect\citeauthoryear{{Kawka} \& {Vennes}}{{Kawka} \&
  {Vennes}}{2011}]{kawka2011}
{Kawka} A.,  {Vennes} S.,  2011, \mn@doi [\aap] {10.1051/0004-6361/201117078},
  \href {https://ui.adsabs.harvard.edu/abs/2011A&A...532A...7K} {532, A7}

\bibitem[\protect\citeauthoryear{{Kawka} \& {Vennes}}{{Kawka} \&
  {Vennes}}{2014}]{kawka2014}
{Kawka} A.,  {Vennes} S.,  2014, \mn@doi [\mnras] {10.1093/mnrasl/slu004},
  \href {https://ui.adsabs.harvard.edu/abs/2014MNRAS.439L..90K} {439, L90}

\bibitem[\protect\citeauthoryear{{Kawka}, {Vennes}, {Schmidt}  \&
  {Koch}}{{Kawka} et~al.}{2007}]{kawka2007}
{Kawka} A.,  {Vennes} S.,  {Schmidt} G.~D.~{Wickramasinghe} D.~T.,   {Koch} R.,
   2007, \mn@doi [\apj] {10.1086/509072}, \href
  {https://ui.adsabs.harvard.edu/abs/2007ApJ...654..499K} {654, 499}

\bibitem[\protect\citeauthoryear{{Kawka}, {Vennes}, {Ferrario}  \&
  {Paunzen}}{{Kawka} et~al.}{2019}]{kawka2019}
{Kawka} A.,  {Vennes} S.,  {Ferrario} L.,   {Paunzen} E.,  2019, \mn@doi
  [\mnras] {10.1093/mnras/sty3048}, \href
  {https://ui.adsabs.harvard.edu/abs/2019MNRAS.482.5201K} {482, 5201}

\bibitem[\protect\citeauthoryear{{Koester}, {G{\"a}nsicke}  \&
  {Farihi}}{{Koester} et~al.}{2014}]{koester2014}
{Koester} D.,  {G{\"a}nsicke} B.~T.,   {Farihi} J.,  2014, \mn@doi [\aap]
  {10.1051/0004-6361/201423691}, \href
  {https://ui.adsabs.harvard.edu/abs/2014A&A...566A..34K} {566, A34}

\bibitem[\protect\citeauthoryear{{Landstreet} \& {Bagnulo}}{{Landstreet} \&
  {Bagnulo}}{2019}]{landstreet2019}
{Landstreet} J.~D.,  {Bagnulo} S.,  2019, \mn@doi [\aap]
  {10.1051/0004-6361/201936009}, \href
  {https://ui.adsabs.harvard.edu/abs/2019A&A...628A...1L} {628, A1}

\bibitem[\protect\citeauthoryear{{Lenz} \& {Breger}}{{Lenz} \&
  {Breger}}{2005}]{lenz2005}
{Lenz} P.,  {Breger} M.,  2005, \mn@doi [Communications in Asteroseismology]
  {10.1553/cia146s53}, \href
  {https://ui.adsabs.harvard.edu/abs/2005CoAst.146...53L} {146, 53}

\bibitem[\protect\citeauthoryear{{Putney}}{{Putney}}{1997}]{putney1997}
{Putney} A.,  1997, \mn@doi [\apjs] {10.1086/313037}, \href
  {https://ui.adsabs.harvard.edu/abs/1997ApJS..112..527P} {112, 527}

\bibitem[\protect\citeauthoryear{{Reid}, {Liebert}  \& {Schmidt}}{{Reid}
  et~al.}{2001}]{reid2001}
{Reid} I.~N.,  {Liebert} J.,   {Schmidt} G.~D.,  2001, \mn@doi [\apjl]
  {10.1086/319481}, \href
  {https://ui.adsabs.harvard.edu/abs/2001ApJ...550L..61R} {550, L61}

\bibitem[\protect\citeauthoryear{{Ricker} et~al.,}{{Ricker}
  et~al.}{2015}]{ricker2015}
{Ricker} G.~R.,  et~al., 2015, \mn@doi [Journal of Astronomical Telescopes,
  Instruments, and Systems] {10.1117/1.JATIS.1.1.014003}, \href
  {https://ui.adsabs.harvard.edu/abs/2015JATIS...1a4003R} {1, 014003}

\bibitem[\protect\citeauthoryear{{Schreiber}, {Belloni}, {G{\"a}nsicke}  \&
  {Parsons}}{{Schreiber} et~al.}{2021}]{schreiber2021}
{Schreiber} M.~R.,  {Belloni} D.,  {G{\"a}nsicke} B.~T.,   {Parsons} S.~G.,
  2021, \mn@doi [\mnras] {10.1093/mnrasl/slab069}, \href
  {https://ui.adsabs.harvard.edu/abs/2021MNRAS.506L..29S} {506, L29}

\bibitem[\protect\citeauthoryear{{Stephan}, {Naoz}, {Gaudi}  \&
  {Salas}}{{Stephan} et~al.}{2020}]{stephan2020}
{Stephan} A.~P.,  {Naoz} S.,  {Gaudi} B.~S.,   {Salas} J.~M.,  2020, \mn@doi
  [\apj] {10.3847/1538-4357/ab5b00}, \href
  {https://ui.adsabs.harvard.edu/abs/2020ApJ...889...45S} {889, 45}

\bibitem[\protect\citeauthoryear{{Subasavage}, {Henry}, {Bergeron}, {Dufour},
  {Hambly}  \& {Beaulieu}}{{Subasavage} et~al.}{2007}]{subasavage2007}
{Subasavage} J.~P.,  {Henry} T.~J.,  {Bergeron} P.,  {Dufour} P.,  {Hambly}
  N.~C.,   {Beaulieu} T.~D.,  2007, \mn@doi [\aj] {10.1086/518739}, \href
  {https://ui.adsabs.harvard.edu/abs/2007AJ....134..252S} {134, 252}

\bibitem[\protect\citeauthoryear{{Swan}, {Farihi}, {Melis}, {Dufour}, {Desch},
  {Koester}  \& {Guo}}{{Swan} et~al.}{2023}]{swan2023}
{Swan} A.,  {Farihi} J.,  {Melis} C.,  {Dufour} P.,  {Desch} S.~J.,  {Koester}
  D.,   {Guo} J.,  2023, \mn@doi [\mnras] {10.1093/mnras/stad2867}, \href
  {https://ui.adsabs.harvard.edu/abs/2023MNRAS.526.3815S} {526, 3815}

\bibitem[\protect\citeauthoryear{{Tout}, {Wickramasinghe}, {Liebert},
  {Ferrario}  \& {Pringle}}{{Tout} et~al.}{2008}]{tout2008}
{Tout} C.~A.,  {Wickramasinghe} D.~T.,  {Liebert} J.,  {Ferrario} L.,
  {Pringle} J.~E.,  2008, \mn@doi [\mnras] {10.1111/j.1365-2966.2008.13291.x},
  \href {https://ui.adsabs.harvard.edu/abs/2008MNRAS.387..897T} {387, 897}

\bibitem[\protect\citeauthoryear{{Uzundag} et~al.,}{{Uzundag}
  et~al.}{2023}]{uzundag2023}
{Uzundag} M.,  et~al., 2023, \mn@doi [\mnras] {10.1093/mnras/stad2776}, \href
  {https://ui.adsabs.harvard.edu/abs/2023MNRAS.526.2846U} {526, 2846}

\bibitem[\protect\citeauthoryear{{Valyavin}, {Wade}, {Bagnulo}, {Szeifert},
  {Landstreet}, {Han}  \& {Burenkov}}{{Valyavin} et~al.}{2008}]{valyavin2008}
{Valyavin} G.,  {Wade} G.~A.,  {Bagnulo} S.,  {Szeifert} T.,  {Landstreet}
  J.~D.,  {Han} I.,   {Burenkov} A.,  2008, \mn@doi [\apj] {10.1086/589234},
  \href {https://ui.adsabs.harvard.edu/abs/2008ApJ...683..466V} {683, 466}

\bibitem[\protect\citeauthoryear{{Wilson}, {Hermes}  \&
  {G{\"a}nsicke}}{{Wilson} et~al.}{2020}]{wilson2020}
{Wilson} D.~J.,  {Hermes} J.~J.,   {G{\"a}nsicke} B.~T.,  2020, \mn@doi [\apjl]
  {10.3847/2041-8213/ab9d7b}, \href
  {https://ui.adsabs.harvard.edu/abs/2020ApJ...897L..31W} {897, L31}

\bibitem[\protect\citeauthoryear{{Zuckerman}, {Koester}, {Reid}  \&
  {H{\"u}nsch}}{{Zuckerman} et~al.}{2003}]{zuckerman2003}
{Zuckerman} B.,  {Koester} D.,  {Reid} I.~N.,   {H{\"u}nsch} M.,  2003, \mn@doi
  [\apj] {10.1086/377492}, \href
  {https://ui.adsabs.harvard.edu/abs/2003ApJ...596..477Z} {596, 477}

\bibitem[\protect\citeauthoryear{{Zuckerman}, {Melis}, {Klein}, {Koester}  \&
  {Jura}}{{Zuckerman} et~al.}{2010}]{zuckerman2010}
{Zuckerman} B.,  {Melis} C.,  {Klein} B.,  {Koester} D.,   {Jura} M.,  2010,
  \mn@doi [\apj] {10.1088/0004-637X/722/1/725}, \href
  {https://ui.adsabs.harvard.edu/abs/2010ApJ...722..725Z} {722, 725}

\makeatother
\end{thebibliography}

%

\bsp    
\label{lastpage}
\end{document}